\documentclass{ws-procs975x65}

\newcommand{\R}{\mathcal{R}}

\begin{document}

\title{Hybrid $f(R)$ theories, local constraints, and cosmic speedup}

\author{S. Capozziello$^{1,2}$, T. Harko$^3$,   T. S. Koivisto$^4$, F. S. N. Lobo$^5$, G. J. Olmo$^6$}
\address{$^1$Dipartimento di Scienze Fisiche, Universit\`{a} di Napoli "Federico II", Napoli, Italy, 
$^2$INFN Sez. di Napoli, Compl. Univ. di Monte S. Angelo, Edificio G, Via Cinthia, I-80126, Napoli, Italy, E-mail: capozzie@na.infn.it }
\address{$^3$Department of Mathematics, University College London, Gower Street, London WC1E 6BT, United Kingdom, E-mail: t.harko@ucl.ac.uk }
\address{$^4$ Institute of Theoretical Astrophysics, University of
  Oslo, P.O.\ Box 1029 Blindern, N-0315 Oslo, Norway,  E-mail: tomi.koivisto@fys.uio.no }
\address{$^5$Centro de Astronomia e Astrof\'{\i}sica da Universidade de Lisboa, Campo Grande, Ed. C8 1749-016 Lisboa, Portugal, E-mail:flobo@cii.fc.ul.pt }
\address{$^6$Departamento de F\'{i}sica Te\'{o}rica and IFIC,
Centro Mixto Universidad de Valencia - CSIC. Universidad de
Valencia, Burjassot-46100, Valencia, Spain, E-mail: gonzalo.olmo@csic.es }

\begin{abstract}
We present an extension of general relativity in which an $f(R)$ term \`{a} la Palatini is added to the usual metric Einstein-Hilbert Lagrangian. Expressing the theory in a dynamically equivalent scalar-tensor form, we show that it can pass the Solar System observational tests even if the scalar field is very light or massless. Applications to cosmology and astrophysics, and some exact solutions are discussed.
\end{abstract}


\bodymatter

\section{Introduction}\label{sec:Intro}
A number of alternative theories of gravity have been proposed in the last decade in relation with the cosmic speedup problem. Among them, $f(R)$ theories have attracted special attention. These theories have been mainly studied in the metric\cite{fRmetric} and in the Palatini\cite{Olmo:2011uz} formalisms. To be in agreement with local tests of gravity, the former case generically requires that the non-linear curvature terms represent a short-range, massive scalar degree of freedom. The latter introduces non-perturbative effects that severely constrain the kind of non-linear terms, ruling out infrared corrections. Here we present a hybrid combination\cite{fX} of both approaches which exhibits interesting new features that allow to avoid the typical shortcomings of the pure metric and Palatini $f(R)$ theories. Recent applications are briefly summarized.

\section{Field equations and weak-field limit.}
The hybrid metric-Palatini action
\begin{equation} \label{eq:S_hybrid}
S= \frac{1}{2\kappa^2}\int d^4 x \sqrt{-g} \left[ R + f(\R)\right] +S_m \ ,
\end{equation}
where $S_m$ is the matter action, $\kappa^2\equiv 8\pi G$, $R$ is the Einstein-Hilbert term,
$\R \equiv  g^{\mu\nu}\R_{\mu\nu} $ is the Palatini curvature, and $\R_{\mu\nu}$ is defined in
terms of an independent connection $\hat{\Gamma}^\alpha_{\mu\nu}$  as
$\R_{\mu\nu} \equiv \hat{\Gamma}^\alpha_{\mu\nu , \alpha} - \hat{\Gamma}^\alpha_{\mu\alpha ,
\nu} + \hat{\Gamma}^\alpha_{\alpha\lambda}\hat{\Gamma}^\lambda_{\mu\nu}
-\hat{\Gamma}^\alpha_{\mu\lambda}\hat{\Gamma}^\lambda_{\alpha\nu}$, admits the following scalar-tensor representation
\begin{equation} \label{eq:S_scalar2}
S=\int \frac{d^4 x \sqrt{-g} }{2\kappa^2}\left[ (1+\phi)R +\frac{3}{2\phi}\partial_\mu \phi
\partial^\mu \phi -V(\phi)\right]+S_m .
\end{equation}
In the weak-field limit and far from the sources, the scalar field behaves as $\phi(r)\approx \phi_0+\frac{2G}{3}\frac{\phi_0 M}{r}e^{-m_\varphi r}$, whereas the metric perturbations $h_{\mu\nu}\approx g_{\mu\nu}-\eta_{\mu\nu}$  take the form
\begin{equation}
h_{00}^{(2)}(r)= \frac{2G_{\rm eff} M}{r} +\frac{V_0}{1+\phi_0}\frac{r^2}{6}  \ , \
h_{ij}^{(2)}(r)= \left(\frac{2\gamma G_{\rm eff} M}{r} -\frac{V_0}{1+\phi_0}\frac{r^2}
{6}\right)\delta_{ij}\ ,
\end{equation}
where $M=\int d^3x \rho(x)$, 
%
\begin{equation}
G_{\rm eff}\equiv  \frac{G}{1+\phi_0}\left[1-\left({\phi_0}/{3}\right)e^{-m_\varphi r}\right],\qquad {\rm and} \qquad \gamma \equiv  \frac{1+\left(\phi_0/3\right)e^{-m_\varphi r}}{1-\left(\phi_0/3\right)e^{-
m_\varphi r}}\,.
\end{equation}
As is clear from these expressions, the coupling of the scalar field to the local system depends on the amplitude of the background value $\phi_0$. If $\phi_0$ is small, then $G_{\rm eff}\approx G$ and $\gamma\approx 1$ regardless of the value of the effective mass $m_\varphi^2$. This contrasts with the result obtained in the metric version of $f(R)$ theories \cite{Olmo:2005jd} and allows to pass the solar system tests even if the scalar field is very light as long as $\phi_0$ is sufficiently small.

\section{Applications}
A light scalar field has a long interaction range. This implies that our theory can modify the long-range gravitational interaction without having any relevant impact on the dynamics of local systems. Thus, the hybrid metric-Palatini theory opens up new possibilities to approach the problems of both dark energy and dark matter. The cosmological applications of this theory have been investigated recently\cite{fX1} by the authors. Criteria to obtain cosmic acceleration were analyzed, and the field equations were formulated  as a dynamical system. Several classes of  cosmological solutions describing both accelerating and decelerating universes were explicitly obtained. Furthermore, the cosmological perturbation equations were derived and applied to uncover the signatures of  these models in the large-scale structure.\\
More recently, we have analysed the generalized virial theorem\cite{Capozziello:2012qt}. Taking into account the relativistic collisionless Boltzmann equation, we  showed that the new geometric terms in the field equations provide an effective contribution to the gravitational potential energy. Indeed, it was shown that the total virial mass is proportional to the effective mass associated with the new terms generated by the effective scalar field, and the baryonic mass. This shows that the mass discrepancy in clusters of galaxies might be due to the effects of modified dynamics. We found that the mass associated to the scalar field and its effects extend beyond the virial radius of the clusters of galaxies. In the context of the galaxy cluster velocity dispersion profiles predicted by the our model, the generalized virial theorem can be an efficient tool in observationally testing the viability of this class of generalized gravity models. Thus, hybrid metric-Palatini gravity provides an effective alternative to the dark matter paradigm of present day cosmology and astrophysics.\\
From a purely theoretical perspective, the general conditions for wormhole solutions according to the null energy condition at the throat in the hybrid $f(R)$ theory have also been considered\cite{fX2}. Several wormhole type solutions were also obtained and analyzed. In the first solution, the redshift function and the scalar field were specified, and the potential was chosen so that the modified Klein-Gordon equation can be simplified. In the second example, by adequately specifying the metric functions and choosing the scalar field, we found an asymptotically flat spacetime.

\section{Conclusions and perspectives}
In a series of works we have shown that a combination of metric and Palatini curvature terms yield a purely geometric theory of gravity in which a light scalar degree of freedom is able to modify the large scale structure without damaging the well-tested dynamics of local systems. The extra terms appearing in the gravitational field equations can describe the early accelerated dynamics of the Universe and explain the matter deficit in galactic clusters. All the relevant astrophysical quantities, including the
``dark mass'' associated to the equivalent scalar-tensor description,  can be expressed in terms of the physical parameters of the model --  the scalar field, its potential, as well as the coupling parameters. Therefore, these results open the possibility of directly testing the hybrid modified $f(R)$ type gravitational models  by using direct astronomical and astrophysical observations. Future work will focus on the dynamics of massive test particles at galactic and extra-galactic scales, and in the formal mathematical aspects of this theory.

\section{Acknowledgments}
SC is supported by INFN (iniziative specifiche NA12 and OG51). TSK is supported by the Research Council of Norway.  FSNL acknowledges financial support of the Funda\c{c}\~{a}o para a Ci\^{e}ncia e Tecnologia through the grants CERN/FP/123615/2011 and CERN/FP/123618/2011. GJO is supported by the Spanish grant FIS2011-29813-C02-02 and the JAE-doc program of the Spanish Research Council (CSIC).

\end{document}